\documentclass[a4paper, 12pt]{article}
\usepackage[utf8]{inputenc}
\usepackage[english]{babel}
\usepackage{amsmath, amssymb, amsfonts}
\usepackage{fullpage, indentfirst, cite}
\usepackage{epsfig, graphicx, graphics}
\usepackage{color}

\newcommand{\be}{\begin{equation}}
\newcommand{\ee}{\end{equation}}

\title{
{\Large Constraining heavy decaying dark matter with the high energy gamma-ray limits}
\date{}
\author{O.~E.~Kalashev and M.~Yu.~Kuznetsov\footnote{mkuzn@inr.ac.ru}
\vspace{.2cm}\\
\footnotesize \it Institute for Nuclear Research of the Russian Academy of Sciences, \\
\footnotesize \it 60th October Anniversary Prospect 7a, 117312 Moscow, Russia}
}

\begin{document}

\begin{flushright}
INR-TH-2016-019
\end{flushright}

{\let\newpage\relax\maketitle}

\begin{abstract}
We consider decaying dark matter with masses $10^{7} \lesssim M \lesssim 10^{16}$ GeV, as
a source of ultra-high energy (UHE) gamma rays. Using recent limits on UHE gamma-ray flux for energies
$E_\gamma > 2 \cdot 10^{14}$ eV, provided by extensive air shower observatories, we put limits on masses and
lifetimes of the dark matter. We also discuss possible dark matter decay origin of tentative 100 PeV photon flux
detected with EAS-MSU experiment.

\end{abstract}


\section{Introduction}

One of the candidates for the role of dark matter are superheavy
particles~\cite{Kuzmin:1997jua, Berezinsky:1997hy} (see also Ref.~\cite{Khlopov:1987bh, Fargion:1995xs}), that we will denote
as $X$ particles. From the point of view of particle physics they
can be incorporated into various theories (see e.g. Ref.~\cite{Kolb:1998ki, Kuzmin:1999zk}
and references therein). In cosmology these particles could be created at some early stages of the Universe
evolution~\cite{Kuzmin:1997jua, Kofman:1994rk, Felder:1998vq, Berezinsky:1997hy, Chung:1998zb, Chung:1998rq, Kolb:1998ki, Kuzmin:1999zk}.
In this paper we consider indirect detection of superheavy dark matter (SHDM).
The parameters that can be experimentally constrained in this approach are mass, annihilation
cross-section and lifetime of the dark matter particles.
While there are several constrains on $X$ particles mass $M_X$ imposed by various scenarios
of the dark matter production~\cite{Kolb:1998ki, Kuzmin:1998kk, Chung:1998zb, Chung:2004nh, Gorbunov:2012ij},
in this study we conservatively consider the full range of $M_X$ accessible for indirect
observation in recent high energy cosmic ray experiments, namely: $10^{7} \lesssim M_X \lesssim 10^{16}$ GeV.
The detection of the annihilation signal of particles with these masses is far beyond reach
of the modern experiments because of the unitarity bound on the $X$ particles annihilation
cross-section~\cite{Aloisio:2006yi}: $\sigma_{ann.} \sim 1/M_{X}^2$.
Therefore, in this work we are focusing on the case of decaying DM with long lifetime $\tau \gg 10^{10}$ yr.

Modern cosmic ray experiments allow to study primary particles composition based on
observed  extensive air showers  (EAS) properties. The spectrum of protons and nuclei with $E>100$ TeV
has been studied in detail in several experiments. In contrast only upper
limits on gamma ray fluxes in the same energy range have been obtained so far~\footnote{We should also mention a
tentative result of primary gamma detection in EAS-MSU experiment~\cite{Fomin:2013mia, Fomin:2014ura}.}.
In this paper we are using these limits to build constraints on decaying SHDM.
For the highest energy ($E \gtrsim 10^{18}$ eV) the recent constraints on gamma
ray flux are given by Pierre Auger Observatory~\cite{Aab:2015bza, Abreu:2011pf},
Telescope Array experiment~\cite{Rubtsov:ICRC2015} and Yakutsk experiment~\cite{Glushkov:2009tn}.
Among the constraints of lower energy gamma flux are the results of KASCADE-Grande~\cite{Kang:2015gpa},
KASCADE~\cite{Schatz:2003} and CASA-MIA~\cite{Chantell:1997gs}.

The main motivation for this study is to refine constraints on SHDM parameters
using all currently available experimental data. For previous works on the same subject
see e.g.~\cite{Aloisio:2006yi, Kalashev:2008dh, Murase:2012xs}.
In recent years the interest to the subject has grown~\cite{Murase:2015gea, Aloisio:2015lva, Esmaili:2015xpa, Dev:2016qbd}
due to PeV neutrino events observation by IceCube~\cite{Aartsen:2013jdh}.

This paper is organized as follows. In Sec.~\ref{flux} we briefly review SHDM decay physics,
consider assumptions about source distribution and propagation of photons in cosmic medium and calculate
photon flux from the decay of SHDM. In Sec.~\ref{results} we compare our results with existing limits on
high energy photon flux and constraint SHDM mass and lifetime.

\section{Gamma-ray flux from SHDM decay}
\label{flux}

The decay of super-heavy particles $X$ was in detail studied in several
works~\cite{Berezinsky:2000up, Sarkar:2001se, Aloisio:2003xj, Barbot:2002ep, Barbot:2002gt}.
 In this work we concentrate on QCD decay channels, since in
this case relatively large flux of photons is produced which
makes them easier to constrain with experimental photon flux limits. 
Note that other decay modes (i.e. leptonic) may also lead to
some photon flux either via direct gamma production or by means
of interactions of products (i.e. electrons) with cosmic microwave
background (CMB) and galactic media. Though these channels may be
important only if QCD decay is relatively suppressed. For a review of
various DM decay modes see Ref.~\cite{Cirelli:2010xx}. 

We consider the two-body decay into quark--antiquark
or gluon-gluon pair. The following QCD cascade develops down in energy until the hadronization
occurs. As a result of hadronization and subsequent decay of unstable hadrons
particles such as protons, photons, electrons and neutrinos are produced.
 It is important to note, that the impact of electroweak interactions
on the hadronic decay channels is subdominant with respect to other
uncertainties of this calculation (e.g. the choice of fragmentation functions, see below).
For low $M_X$ --- low energy region we validate this assumption
comparing the decay spectra of Refs.~\cite{Cirelli:2010xx, Ciafaloni:2010ti}
with and without EW corrections. For high $M_X$ and high energies we
compare the spectra obtained in Ref.~\cite{Barbot:2002ep, Barbot:2002gt},
where full MSSM was considered, with that of Ref.~\cite{Aloisio:2003xj},
where authors considered only SUSY QCD interactions. In both energy regions
the difference was found negligible. 

In some earlier works (see e.g. Ref.~\cite{Kalashev:2008dh}) the observed 
shape of proton spectrum was also used to constrain the SHDM parameters.
This method gives weaker results than the usage of $\gamma$--limits since
the proton flux is dominated by the particles of astrophysical origin. Therefore
in this study we do not consider proton flux from SHDM decay.

\begin{figure}
   \includegraphics[width=13.50cm]{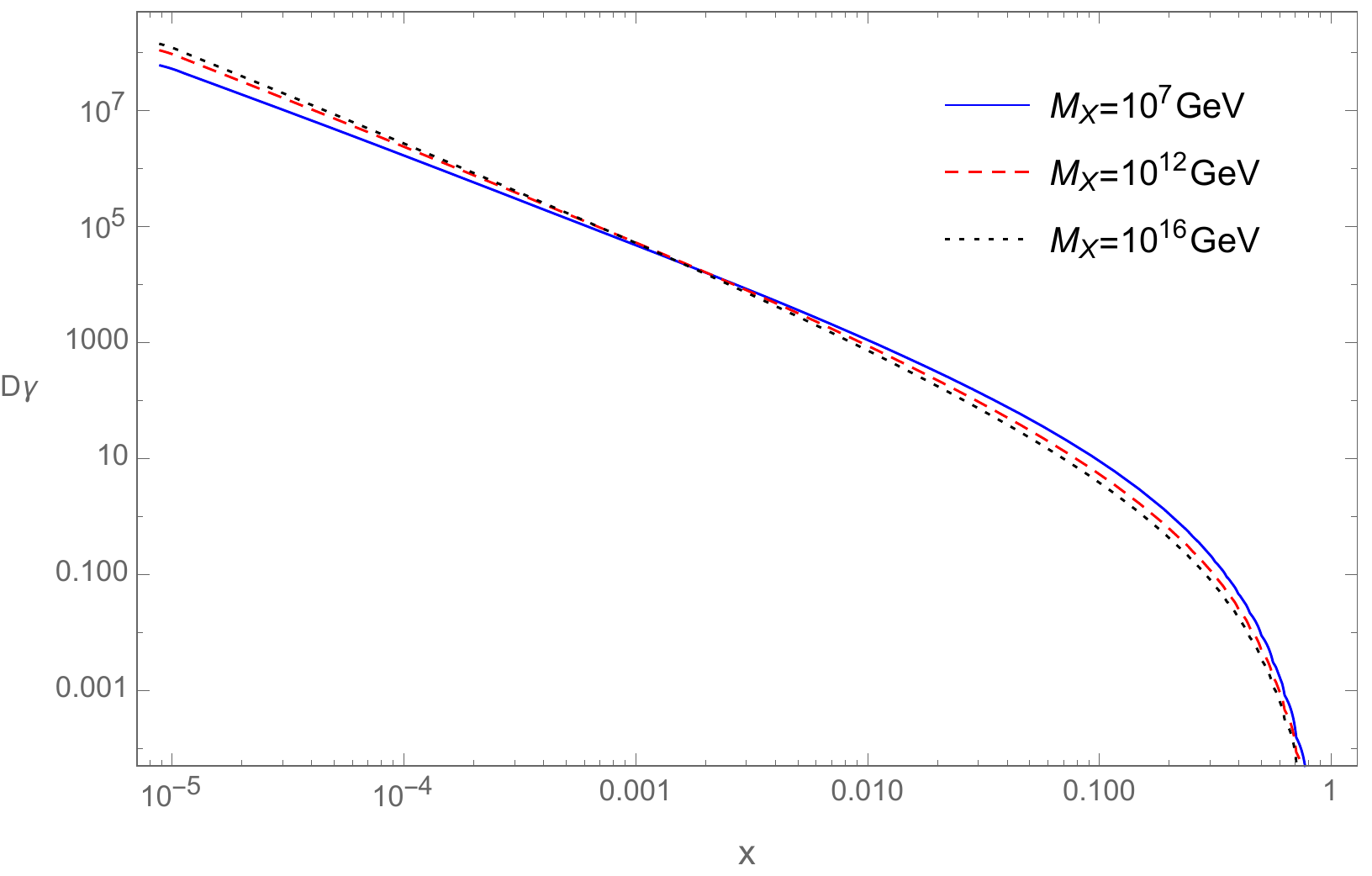}
   \caption{Prompt photon spectra of $X$--particle decay.}
   \label{prompt_spectra}
\end{figure}

Technically the spectra of the $X$-particle decay is defined similarly to the spectra of
$e^+ e^- \rightarrow hadrons$ process~\cite{Hirai:2007cx}:
\be
F^h (x,s) = \sum_i
\int\limits_x^1 \frac{dz}{z} C_i(z,\alpha_s(s)) D_i^h(\frac xz,s) 
\ee
where $x~\equiv~\frac{2 \cdot E}{M_X}$ is the energy of hadron as a fraction of the total
available energy, $D^h_i(x,s)$ are the fragmentation functions of hadron of the type $h$
from the parton of the type $i$, $C_i(z,\alpha_s(s))$ are the coefficient functions and
the summation goes over all types of partons $i= \{u,\bar{u},d,\bar{d},...,g\}$.
The normalization to the $X$ particle decay width is assumed. 
For the leading order in $\alpha_s$ the coefficient functions $C_i$ are proportional to $\delta(1-z)$
and the total spectrum is equal to the sum of fragmentation functions $F^h(x,s) = \sum_i D^h_i(x,s)$.
Given the fragmentation function at some scale $s$ we can
evolve it to another scale using DGLAP equations~\cite{GLD, AP}:
\be
\frac{\partial D_i^h(x,s)}{\partial \ln s} = \sum_j \frac{\alpha_s(s)}{2\pi}P_{ij}(x,\alpha_s(s)) \otimes
D_j^h(x,s)\,,
\ee
where $\otimes$ denotes the convolution $f(x) \otimes g(x) \equiv \int_x^1 dz/z f(z)g(x/z)=\\
\int_x^1 dz/z f(x/z)g(z)$ and $P_{ij}(x,s)$ is the splitting function for the parton branching $i \rightarrow j$.
Since we study the process on the scale  $M_X \gg m_q$  we assume all $N_f$ quark flavors are coupled to gluon similarly
and we can confine ourselves to considering only the mixing of gluon fragmentation function with a quark
singlet fragmentation function: 
\be
D_q^h(x,s)=\frac{1}{N_f} \sum\limits_{i=1}^{N_f} [D_{q_i}^h(x,s) + D_{\bar{q_i}}^h(x,s)]\, .
\ee
Then DGLAP equations take the form:
\be
 \frac{\partial}{\partial \ln s}\left(\begin{array}{c}
  D_q^h(x,s)\\
  D_g^h(x,s)\\
                  \end{array} \right)=
\left( \begin{array}{cc}
       P_{qq}(x,s) &  P_{gq}(x,s) \\
       2N_f P_{qg}(x,s) & P_{gg}(x,s) \\
\end{array} \right)
 \otimes
 \left( \begin{array}{c}
  D_q^h(x,s)\\
  D_g^h(x,s)\\
\end{array} \right)\, .
\ee
In this study we use the code kindly provided by the authors of Ref.~\cite{Aloisio:2003xj}.
This code evaluates the DGLAP equations numerically in the leading order in $\alpha(s)$.
We use the initial fragmentation functions from the Ref.~\cite{Hirai:2007cx}
parametrized on the scale $M_Z$ and extrapolated to the region $10^{-5} \le x \le 1$.
Although the low $x$ tail is unreliable at this scale, the results obtained for the high scales $M_X$
agree with that obtained by Monte-Carlo simulation, as it was shown in~\cite{Aloisio:2003xj}.
Fortunately, the spectra calculated in this region of $x$ are enough to constrain the results
with the experiment in the mass range of interest: $10^{7} \le M_X \le 10^{16}$ GeV.
In this paper we calculate only prompt photon spectra of $\pi^0$s decay and neglect the smaller
amount of photons from inverse Compton scattering (ICS) of prompt $e^\pm$ on the interstellar
background photons.  While for the leptonic decay channels the relative contribution of inverse
Compton photons to the full spectrum can be significant~\cite{Esmaili:2015xpa},
for hadronic channels it is at least by order of magnitude lower~\cite{Cirelli:2010xx},
so we neglect the contribution from prompt $e^\pm$ via ICS in this study. 
Following~\cite{Aloisio:2003xj} we also neglect
roughly $10\%$ contribution of other mesons decay.
Then the photon spectrum of the $X$--particle decay is given by:
\be
D^\gamma(x) = 2 \int\limits_x^1 \frac{dz}{z} \: D^{\pi^0}(z) \,,
\ee
where $D^{\pi^0}(x,s) \equiv [D^{\pi^0}_q(x,s) + D^{\pi^0}_g(x,s)]$.
The examples of prompt photon spectra for decay of $X$--particle
with different masses are shown in Fig~\ref{prompt_spectra}.

Having the injected photon spectra we can calculate the corresponding
photon flux reaching the Earth. We use the following assumptions.
First of all we neglect the flux coming from extragalactic region.
Starting at $E_{\gamma} = 2 \cdot 10^{14}$ eV, which is the lowest energy where
the EAS experiments provide photon limits, and up to $E_{\gamma} \simeq 2 \cdot 10^{18}$ eV
the photon attenuation length doesn't exceed the size of our Galaxy halo.
Then, up to the highest experimentally tested energy $E_{\gamma} = 10^{20}$ eV
photons can come from a region of size not exceeding $50$ Mpc, which contribution to
the flux is about $1\%$ of that from our Galaxy~\cite{Dubovsky:1998pu}.

\begin{figure}
   \includegraphics[width=13.50cm]{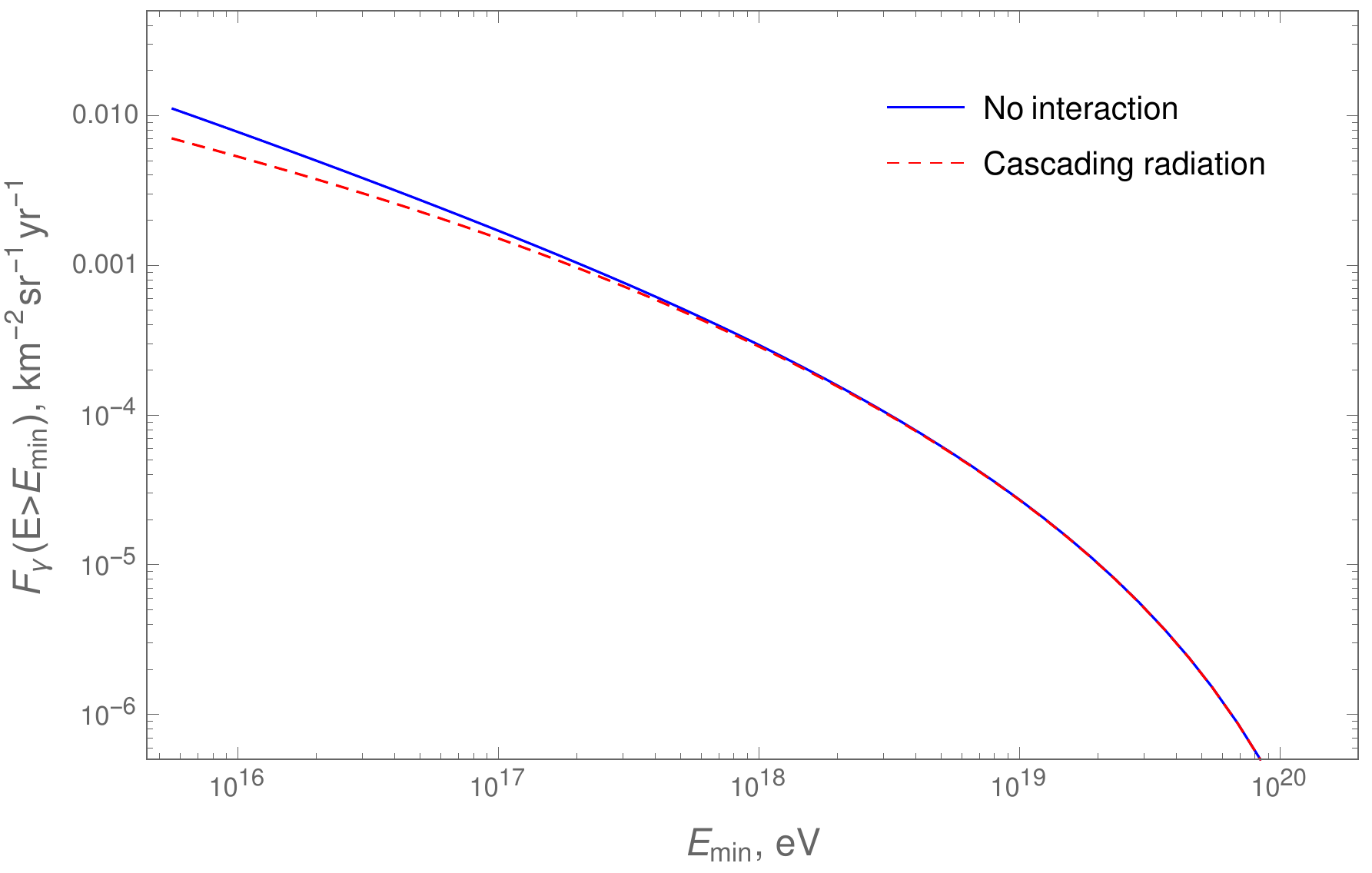}
   \caption{Integral photon flux from SHDM decay in our Galaxy halo as
   received by Telescope array experiment, $M_X=10^{12}$ GeV, 
   without interactions of photons with medium (solid line),  with photon interactions with CMB and secondary cascade radiation
   included (dashed curve)}
   \label{cascaded_flux}
\end{figure}

For the galactic photon flux calculation we use Navarro-Frenk-White dark matter
distribution~\cite{Navarro:1995iw, Navarro:1996gj} with the parametrization for  Milky Way 
from  Ref.~\cite{Cirelli:2010xx}~\footnote{ For comparison we have also tested Burkert dark matter
profile~\cite{Burkert:1995yz}. }. 
We assume photons being radiated isotropically in the decay
of $X$--particle. As it was mentioned above, for photons with $E \gtrsim 10^{18}$ eV
the attenuation length in interstellar medium exceeds the size of our Galaxy halo.
This implies that for higher energy photon we can neglect the absorption and cascaded
radiation. Indeed, the comparison of the non-interacting and cascading $\gamma$ fluxes from
our Galaxy (the latter was calculated using the numerical code from Ref.~\cite{Kalashev:2014xna},  see below )
shows that the discrepancy does not exceeds few percent for  $E = 10^{18}$ eV  (see Fig.~\ref{cascaded_flux}).
Therefore we neglect the photon interaction with the medium for $M_X \gtrsim 10^{14}$ GeV.
Using the above assumptions we obtain the following expression for the integral photon flux
received by a given cosmic ray observatory:
\begin{equation}
\label{int_flux}
F(E>E_{min}) = \frac{N(E>E_{min})}{4\pi M_X \tau} \cdot
\frac{\int\limits_{V} \frac{\rho(R) \omega(\delta, a_0, \theta_{max})}{r^2} d V}{ 2\pi  \int\limits_{-\frac{\pi}{2}}^{\frac{\pi}{2}} \omega(\delta, a_0, \theta_{max}) \cos(\delta) d \delta}\; ;
\end{equation}
where $\rho(R)$ --- is a DM density as a function of distance $R$ from galactic center,
$r$ --- is a distance from Earth, $\omega$ --- is a relative exposure of the given
observatory and $N(E>E_{min})$ --- is an integral number
of photons with energies higher than $E_{min}$ produced
in the decay of $X$--particle. Integration in the numerator
takes over all volume of halo  ($R_{max} = 260$ kpc) 
and in the denominator over all sky (the averaging over
right ascension is included in the definition of $\omega$).
The relative exposure $\omega$ is a function of
declination $\delta$, geographical latitude
of the given experiment $a_0$ and the maximal
zenith angle $\theta_{max}$ of particles allowed
for observation in this experiment
(see Refs.~\cite{Sommers:2000us, Aab:2014ila} for details).

For $M_X \lesssim 10^{14}$ GeV we also take into account the
attenuation of photons on CMB using the numerical code~\cite{Kalashev:2014xna}.
The code simulates development of electron-photon cascades on CMB driven by
the chain of $e^\pm$ pair production and inverse Compton scattering. Although
the code allows to calculate the flux of the cascade photons it doesn't take
into account deflections of $e^\pm$ by the halo magnetic field.
Since electrons in the code propagate rectilinearly they produce
less cascade photons. Therefore the calculated flux of photons
should be considered as conservative lower bound.  The propagation
code~\cite{Kalashev:2014xna} also includes attenuation of photons
on extragalactic background light (EBL), though the effect of
EBL is negligible on distances which we consider. 

\section{Comparison with photon limits}
\label{results}

Finally we compare the predicted SHDM signal with the existing experimental upper-limits
on photon flux. For the highest observable cosmic ray energies ($E_{CR} \gtrsim 10^{18}$ eV)
the recent constraints are provided by Pierre Auger Observatory~\cite{Aab:2015bza, Abreu:2011pf},
Telescope Array experiment~\cite{Rubtsov:ICRC2015}  and Yakutsk experiment~\cite{Glushkov:2009tn}, 
while for the lower energies we use the results of  CASA-MIA~\cite{Chantell:1997gs},
KASCADE~\cite{Schatz:2003}, KASCADE-Grande~\cite{Kang:2015gpa} and EAS-MSU~\cite{Fomin:2014ura}. 
For a review of experimental results see e.g. Ref.~\cite{Karg:2015gxa} and references therein.
We should note, that the higher energy limits are more effective for constraining SHDM
since its decay spectra is quite hard, i.e. SHDM photon flux grows slower than the experimental
limits with the decreasing of energy.

\begin{figure}
   \includegraphics[width=13.50cm]{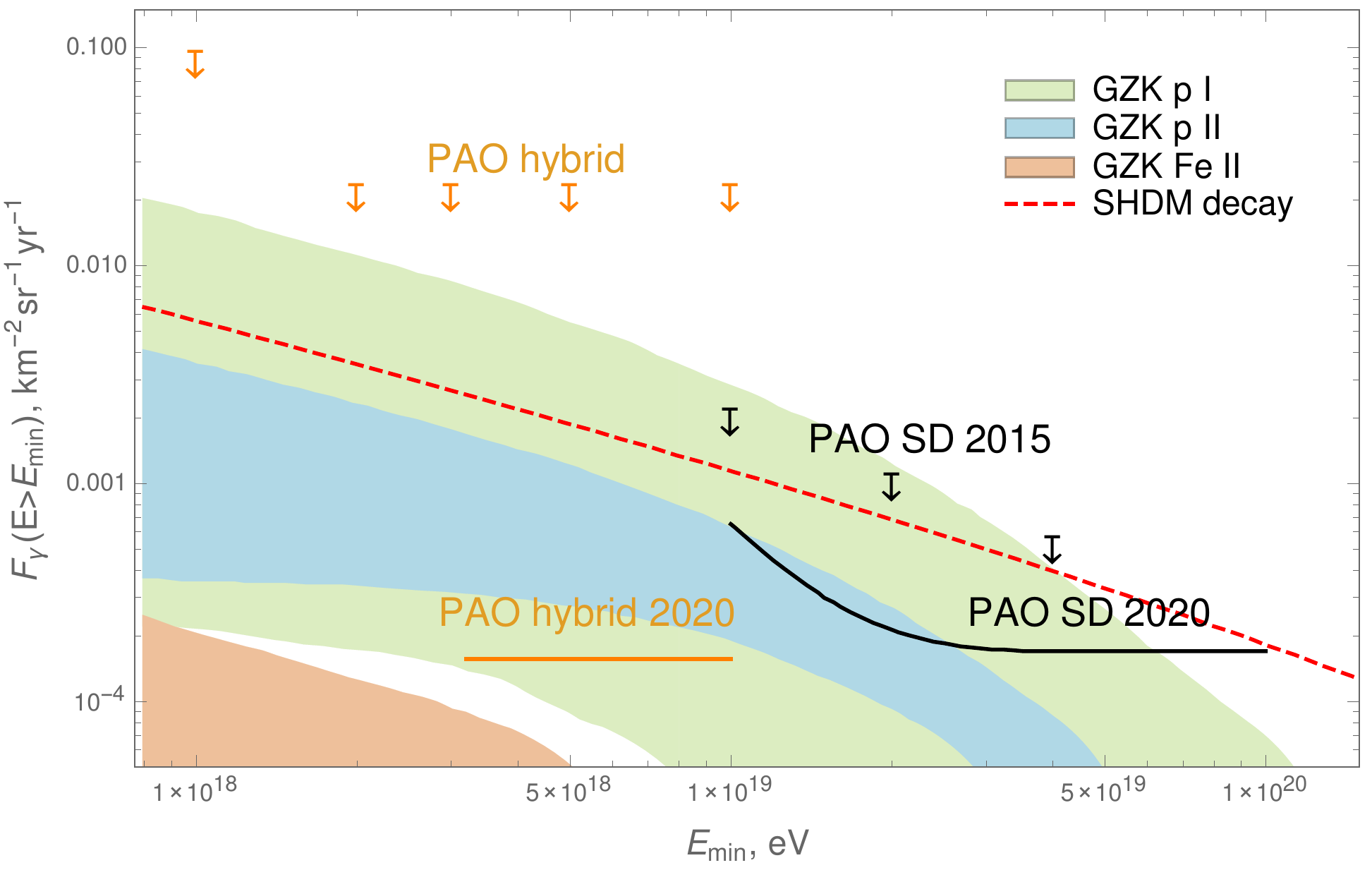}
   \caption{Predicted integral photon flux from decay of SHDM with mass $M_X=10^{14}$ GeV and
    lifetime  $\tau= 2 \cdot 10^{22}$ yr  compared with upper--limits of Pierre Auger
    Observatory~\cite{Aab:2015bza, Abreu:2011pf} and its estimated sensitivity for 2020
    (assuming the upgrade of facility)~\cite{Karg:2015gxa}. Estimates of
    $\gamma$--ray background produced  by attenuation of UHE protons~\cite{Gelmini:2005wu} (green shaded) and
    UHE protons and iron induced cascades~\cite{Hooper:2010ze} (blue and orange shaded) are shown with their
    theoretical uncertainties.}
   \label{flux_limits_hi}
\end{figure}

\begin{figure}
   \includegraphics[width=13.50cm]{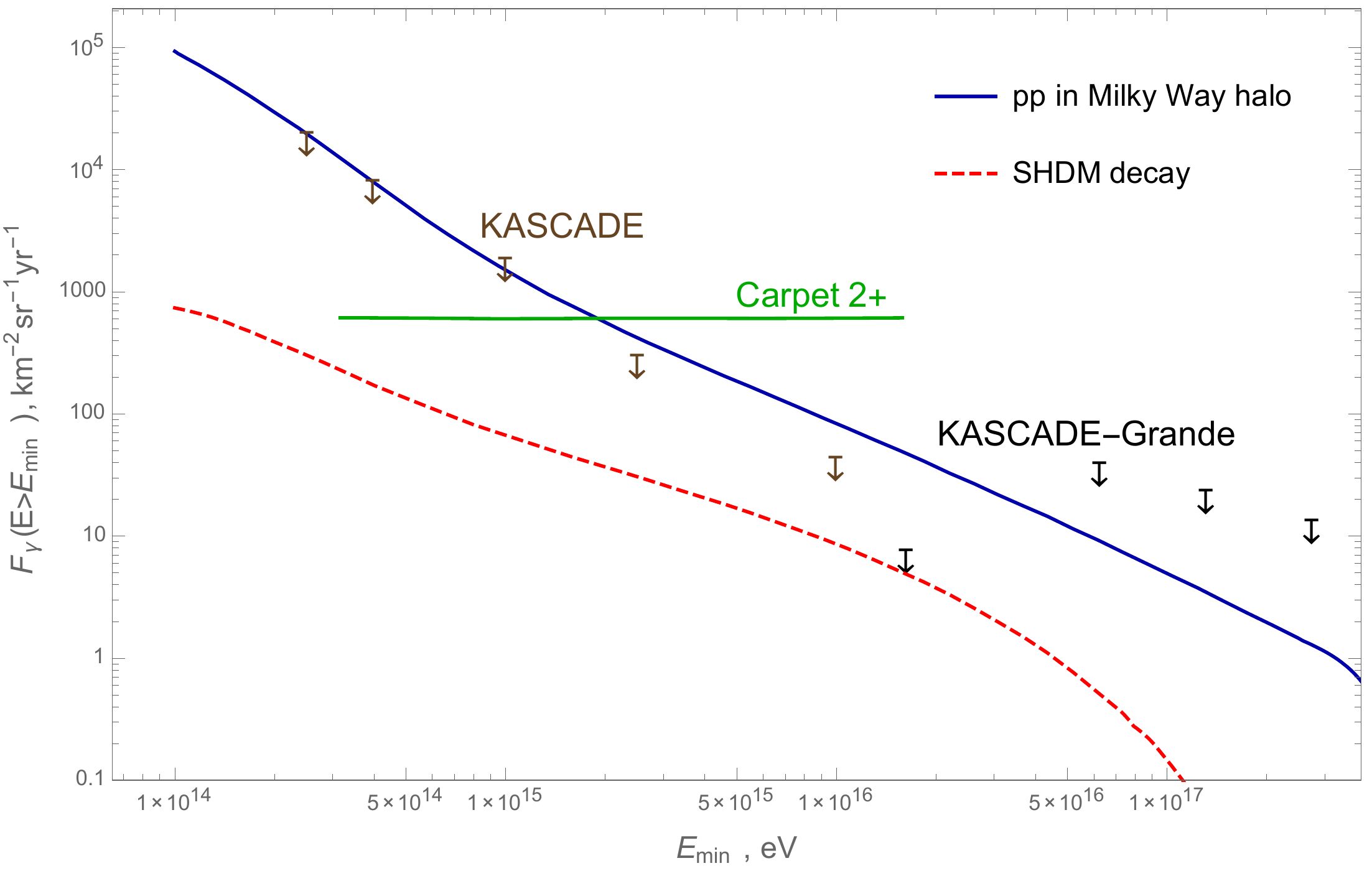}
   \caption{Predicted integral photon flux from decay of SHDM with mass $M_X=10^{9}$ GeV and lifetime  $\tau= 3 \cdot 10^{21}$ yr 
    compared with upper limits of KASCADE and KASCADE-Grande experiments~\cite{Schatz:2003, Kang:2015gpa},
    estimated sensitivity of  Carpet $2+$  experiment~\cite{Dzhappuev:2015hxl}
    and the estimate~\cite{Kalashev:2014vra} of $\gamma$ background from $pp$--interactions in halo.}
   \label{flux_limits_low}
\end{figure}

Another possible contribution to UHE photon flux comes from astrophysics.
UHE protons and nuclei produced by extragalactic sources interact
with CMB and other interstellar background producing secondary electron-photon cascades and neutrinos.
The essentially isotropic flux of photons of this origin has been estimated in several scenarios including proton and nuclei
emitting sources (see e.g. Refs.~\cite{Gelmini:2005wu, Hooper:2010ze, Kalashev:2014vra, Joshi:2013aua, Ahlers:2013xia}).
Contrary to the astrophysical signal the SHDM contribution is anisotropic with maximum flux arriving from the center of Milky Way.
In Figs.~\ref{flux_limits_hi}--\ref{flux_limits_low} we show the $\gamma$--ray
flux limits by KASCADE, KASCADE-Grande and Pierre Auger Observatory
together with predicted SHDM decay photon flux (for certain parameters of SHDM)
and some estimates of astrophysical photon flux.
Also we show the estimated future sensitivity of 
Carpet experiment~\cite{Dzhappuev:2015hxl} in Fig.~\ref{flux_limits_low} and upgraded
PAO~\cite{Karg:2015gxa} in Fig.~\ref{flux_limits_hi}\footnote{Because of the strong anisotropy
of the predicted SHDM signal, we do not show all the existing experimental limits on single
picture. KASCADE and Carpet experiments have approximately the same geographical latitude.}.

\begin{figure}
   \includegraphics[width=13.50cm]{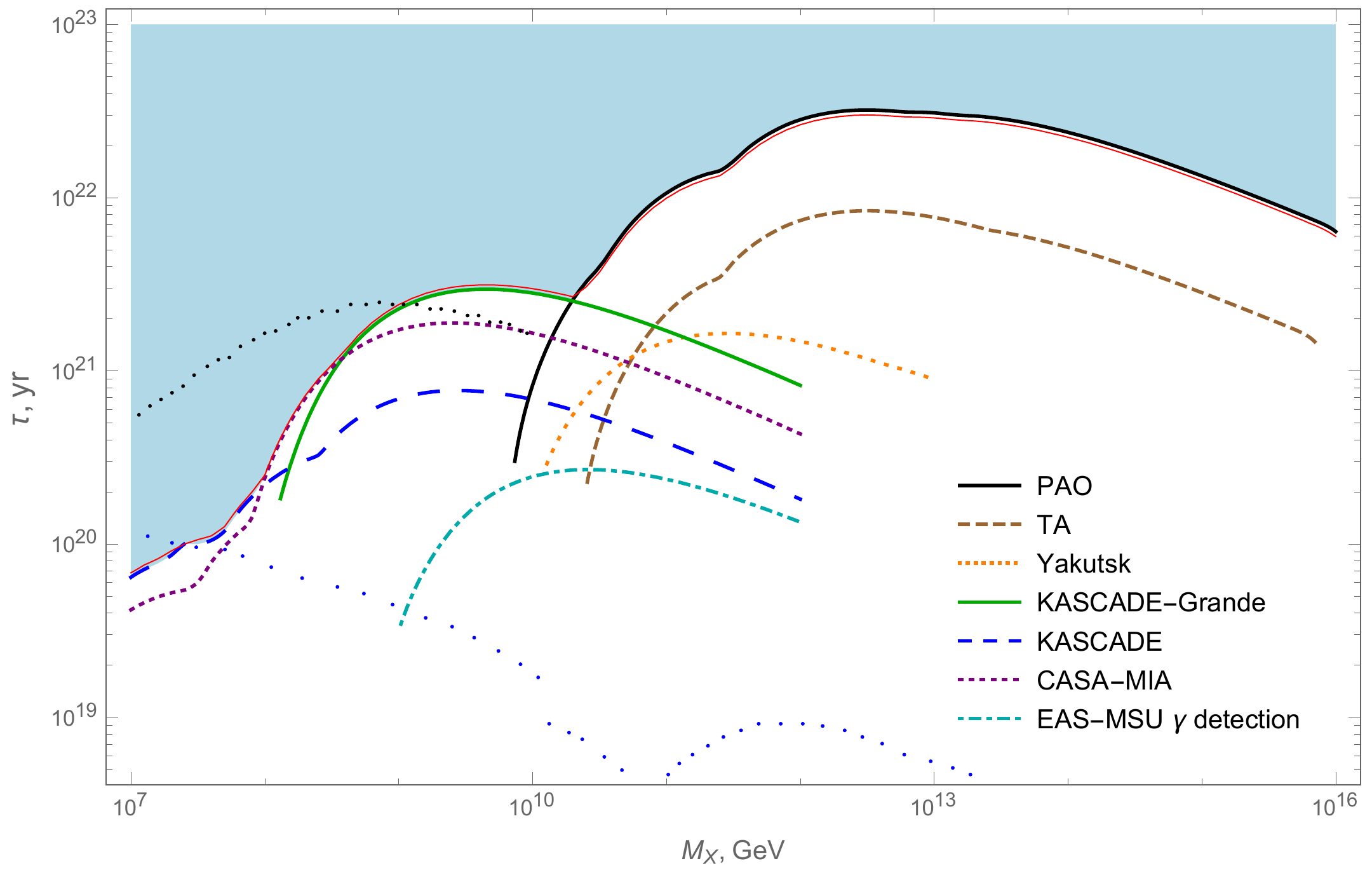}
   \caption{Constraints on mass $M_X$ and lifetime $\tau$ of super heavy dark matter. White area is excluded.
    For comparison we present the constraints obtained with Burkert DM profile (solid thin red line). We also
   show the constraint obtained with neutrino limits: for $X \rightarrow \nu\bar{\nu}$ channel~\cite{Esmaili:2012us} (blue dots)
   and for $X \rightarrow b\bar{b}$ channel~\cite{Murase:2012xs} (black dots).}
   \label{exclusion_plot}
\end{figure}

We compare the constraints of various experiments on SHDM mass and lifetime in Fig.~\ref{exclusion_plot}.
The constraints are built by scanning SHDM parameter space and matching the predicted photon signal with the limits
of the given experiment. The model is considered as excluded as soon as the
signal touches the limit points from below. For EAS-MSU result of photon
detection~\cite{Fomin:2014ura} we show the fit assuming the whole photon flux being produced by SHDM decay.
The constraints based on Pierre Auger Observatory limits are the strongest since this
experiment has largest exposure among UHECR experiments and it is located in the Southern hemisphere
where higher $\gamma$--ray flux coming from galactic center could be detected.
The strongest constraint over all mass range is  $\tau \gtrsim 3 \cdot 10^{22}$ yr at $M_X \simeq 3 \cdot 10^{12}$ GeV. 
It  slightly  improves the result of Ref.~\cite{Aloisio:2015lva} for which the old PAO limits were used.
In the low energy region the best constraints are derived from KASCADE, CASA-MIA and KASCADE-Grande:
minimal lifetime increases from  $\tau \simeq 6 \cdot 10^{19}$ yr at $M_X = 10^7$ GeV to $\tau \simeq 3 \cdot 10^{21}$ yr
at $M_X = 5 \cdot 10^{9}$ GeV  being of the same order as the constraints of
Refs.~\cite{Murase:2012xs, Murase:2015gea, Esmaili:2015xpa} that were obtained in a wider theoretical
context.  The constraints obtained with Burkert dark matter profile is slightly weaker than that
of NFW in the high energy region, where PAO observes the Galactic center,
and stronger for low energies, where constraints are put by Northern hemisphere experiments.

It is also interesting to compare our constraints with those obtained from neutrino limits.
In Ref.~\cite{Murase:2012xs} the neutrino constraints on $\tau$ was imposed for $M_X < 10^{10}$ GeV
and for various decay channels. Our constraints are of the same order as these for $M_X \gtrsim 10^9$ GeV 
but become weaker for $M_X \lesssim 10^9$ GeV. The case of direct decay of
dark matter into neutrino was studied in Ref.~\cite{Esmaili:2012us} for a wide region
$10 < M_X < 10^{19}$ GeV. The constraints on $\tau$ obtained there are of the same order
as ours for $M_X \lesssim 10^8$ GeV and weaker for all higher masses. 

Our constraints have implication for the EAS-MSU tentative result of 100 PeV gamma detection~\cite{Fomin:2014ura}.
We may see that the curve interpreting it as the product of SHDM decay lies deep in the
parameter area excluded by the other experiments, this implies that SHDM component
in EAS-MSU photon signal can not be dominant. 

Discussing these results we may note that although the recent experimental limits
touch the astrophysically predicted region, due to large uncertainty of astrophysical
$\gamma$--ray flux, one can not yet exclude the dominant contribution of SHDM decay.
Nevertheless, one might use the guaranteed i.e. minimal predicted astrophysical gamma flux to
constrain SHDM parameters even stronger. Finally, if $\gamma$--rays are detected the
discrimination between the astrophysical and the SHDM (or other exotic) origin scenario
could be in principal made by analysing the flux anisotropy and energy spectrum.

\section*{Acknowledgements}
We thank S.~Troitsky and G.~Rubtsov for helpful discussions. We are especially indebted to 
R.~Aloisio, V.~Berezinsky and M.~Kachelriess for providing the numerical code solving DGLAP equations.  
This work has been supported by Russian Science Foundation grant 14-22-00161.
 Numerical simulations have been performed in part at the computer cluster of
the Theoretical Physics Department of the Institute for Nuclear Research
of the Russian Academy of Sciences.

\suppressfloats

\end{document}